\documentclass[fleqn,10pt]{wlscirep}
\usepackage[utf8]{inputenc}
\usepackage[T1]{fontenc}
\usepackage{physics}
\usepackage{algorithm, algpseudocode}
\usepackage{xcolor}
\usepackage{diagbox}

\definecolor{GoldYellow}{HTML}{FFC107}

\usepackage{nameref}
\usepackage{hyperref}
\usepackage{cleveref} 

\usepackage[font=footnotesize, labelfont=bf, justification=justified]{caption}
\usepackage{subcaption}
\usepackage{tabularx}
\newcolumntype{Y}{>{\centering\arraybackslash}X}

\title{Quantum Cognition Machine Learning for Forecasting Chromosomal Instability}

\author[1,2,*]{Giuseppe Di Caro}
\author[1,*]{Vahagn Kirakosyan}
\author[1,3]{Alexander G. Abanov}
\author[1,8]{\authorcr Jerome R. Busemeyer}
\author[1,4]{Luca Candelori}
\author[2]{Nadine Hartmann}
\author[2]{\authorcr Ernest T. Lam}
\author[1]{Kharen Musaelian}
\author[1]{Ryan Samson}
\author[1,9]{Harold Steinacker}
\author[1,7]{\authorcr Dario Villani}
\author[1,5]{Martin T. Wells}
\author[2]{Richard J. Wenstrup}
\author[6]{Mengjia Xu}

\affil[1]{Qognitive, Inc., Miami Beach, FL, USA}
\affil[2]{Epic Sciences, San Diego, CA, USA}
\affil[3]{Stony Brook University, Department of Physics and Astronomy, Stony Brook, NY, USA}
\affil[4]{Wayne State University, Department of Mathematics, Detroit, MI, USA}
\affil[5]{Cornell University, Department of Statistics and Data Science, Ithaca, NY, USA}
\affil[6]{New Jersey Institute of Technology, Department of Data Science, Newark, NJ, USA}
\affil[7]{King’s College London, Department of Mathematics, London, UK}
\affil[8]{Indiana University, Department of Psychological and Brain Sciences, Bloomington, IN, USA}
\affil[9]{University of Vienna, Department of Physics, Vienna, Austria}
\affil[*]{Corresponding authors (\href{mailto:giuseppe.dicaro@qognitive.io}{giuseppe.dicaro@qognitive.io}, \href{mailto:vahagn.kirakosyan@qognitive.io}{vahagn.kirakosyan@qognitive.io})}


\begin{abstract}
The accurate prediction of chromosomal instability from the morphology of circulating tumor cells (CTCs) enables real-time detection of CTCs with high metastatic potential in the context of liquid biopsy diagnostics. However, it presents a significant challenge due to the high dimensionality and complexity of single-cell digital pathology data. Here, we introduce the application of Quantum Cognition Machine Learning (QCML), a quantum-inspired computational framework, to estimate morphology-predicted chromosomal instability in CTCs from patients with metastatic breast cancer. QCML leverages quantum mechanical principles to represent data as state vectors in a Hilbert space, enabling context-aware feature modeling, dimensionality reduction, and enhanced generalization without requiring curated feature selection. QCML achieved the highest mean balanced accuracy among the evaluated regression models, while QCML POVM achieved the highest mean ROC--AUC among the classification models. These preliminary findings support QCML as a competitive machine-learning approach for high-dimensional, low-sample-size biomedical applications. QCML enables the simulation of cognition-like learning for the identification of biologically meaningful prediction of chromosomal instability from CTC morphology, offering a novel tool for CTC classification in liquid biopsy.
\end{abstract}

\begin{document}

\flushbottom
\maketitle
\thispagestyle{empty}

\section{Background}
\label{sec:background}

Unlike traditional tissue tests\cite{RN31, RN30}, cell-based liquid biopsy assays enable selection of individual CTCs for the analysis of chromosomal instability using next-generation sequencing by quantification of large-scale state transitions (LST)~\cite{RN1,RN2,RN15,RN17,RN35,RN53,RN52}. Chromosomal instability is a genomic characteristic of cancer cells that drives tumor evolution and metastatic potential~\cite{RN6,RN8,RN9,RN14,RN16,RN23,RN24,RN26,RN34,RN49}. However, whole genome sequencing assays are laborious, requiring a complex workflow that invariably results in a considerable turnaround time that sometimes is not compatible with clinical practice \cite{RN54}. A previous study has shown that we can partially predict chromosomal instability in individual cells by developing algorithms that analyze a range of features, including cell shape, size, morphology, and protein levels, from images of CTCs using an automated digital pathology pipeline \cite{RN1}. Predicting chromosomal instability through morphology offers significant advantages; it can significantly reduce turnaround times compared to whole-genome assays, providing crucial information about the genomic characteristics of CTCs in a patient in a shorter timeframe~\cite{RN1}. Timely information on the presence of CTCs with the highest metastatic potential may be critical for making optimal clinical decisions.

\begin{figure}
  \centering
  \includegraphics[width=\textwidth]{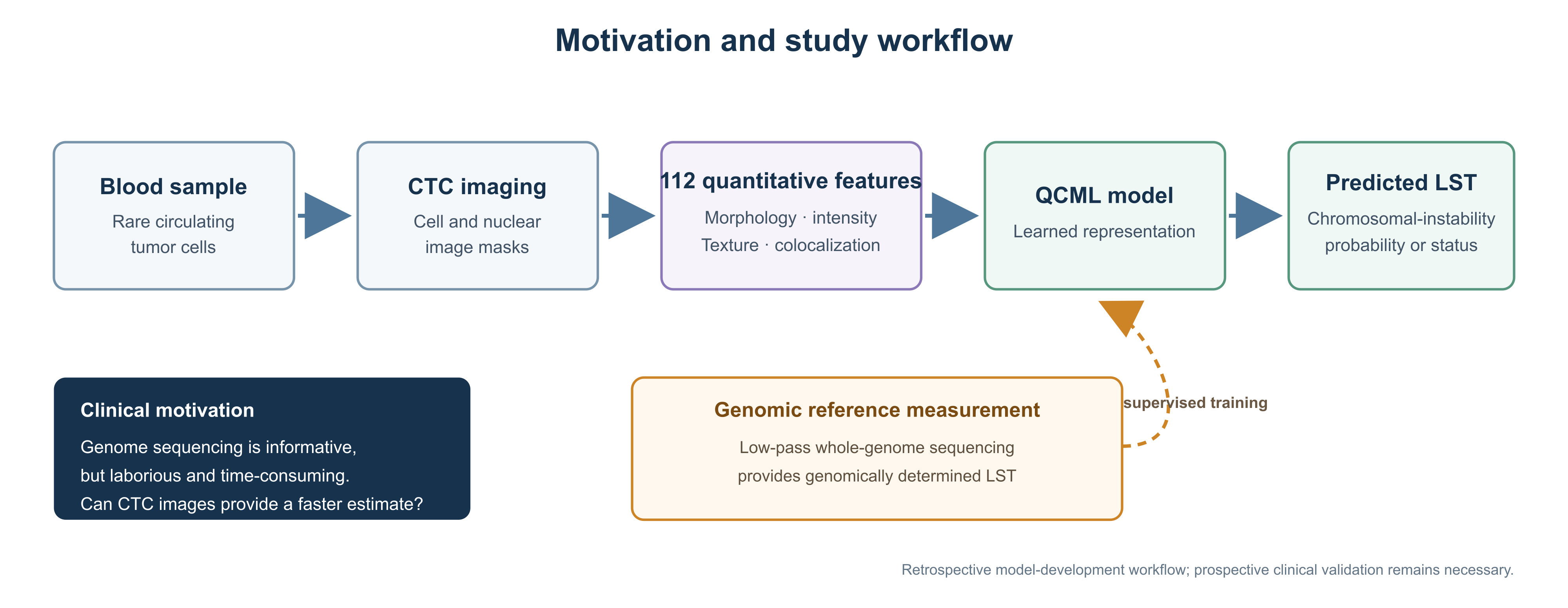}
  \caption{Motivation and study workflow}
  \label{fig:QCML_workflow}
\end{figure}

A key challenge in predicting chromosomal instability through morphology is the utilization of a machine-learning method that accurately classifies morphology patterns from all CTC features and generalizes reproducibly, compatible with potential validation for clinical use~\cite{RN27,RN29,RN55,RN3}. \autoref{fig:QCML_workflow} describes the main motivation and workflow for our study. Key limitations of commonly used machine learning techniques in biology applications, such as support vector machines (SVMs) with Gaussian kernels, include the following \cite{RN27,RN29,RN55,RN3}: 1) The increase in dimensionality that arises from combinations of multiple features exponentially complicates the prediction task, as often seen with cell morphologies. 2) SVMs struggle when classes significantly overlap or when there is label noise, resulting in support vectors that distort the generalization, leading to high misclassification rates and overfitting on independent datasets. 3) The decision boundary learned by a nonlinear SVM is often not biologically interpretable. The biological explainability of the underlying models is crucial to enhancing reproducibility \cite{RN27,RN55,RN29,RN3}.

Quantum Cognition Machine Learning (QCML)~\cite{CandeloriEtAl, MusaelianEtAl, SamsonEtAl, BlackrockBondSimilarity} is an emerging field that introduces a novel approach to machine learning, grounded in the mathematical principles of quantum theory. In QCML, data points are represented as quantum states in a complex Hilbert space, while features and target variables are modeled as Hermitian operators or ``observables''. The ``observables'' are learned by optimizing a particular objective function over the observables’ parameters. Although QCML models use similar objective functions and evaluation metrics as classical machine learning models, they differ fundamentally in how data is represented and how functional dependencies among features are parameterized. QCML models are versatile, capable of handling numerical and categorical data, as well as missing and/or noisy data. They create a global quantum manifold model (in the sense of quantum geometry \cite{Steinacker_book}) of the original data manifold, that is robust to noise and able to generalize well beyond training samples~\cite{CandeloriEtAl, SamsonEtAl, BlackrockBondSimilarity}. Part of this adeptness at controlling variance stems from the fact that the number of parameters of a QCML model scales linearly with the number of features, thus achieving logarithmic economy of representation. For the first time we introduce the QCML Positive Operator-Valued Measure (POVM). QCML POVM is an extension of QCML that allows one to forecast the probability density function of a target, as opposed to point estimate. It naturally lends itself to our problem setup with pLST forecasting where we can produce both real-valued forecasts of pLST (based on expected mean/median) and probability forecasts of pLST being above/below a certain threshold. 

We hypothesize that QCML's ability to generalize gives it an advantage in computing morphology-predicted pLST compared to typical machine-learning algorithms such as SVMs. Here, we test whether QCML may help prevent overfitting and improve prediction performance and reproducibility in metastatic breast cancer cells. Among the regression models, QCML achieved the highest mean balanced accuracy and specificity. Among the classification models, QCML POVM achieved the highest mean ROC--AUC, a threshold-independent performance metric.

\section{Methods}
\label{sec:methods}

\subsection{Data Acquisition and Preparation}

As previously published \cite{RN2}, the CTC assay for metastatic breast cancer follows a non-enrichment strategy where all nucleated cells from patient blood are deposited onto slides and stained using immunofluorescence. We used previously published high-resolution digital image analysis technology of CTCs for metastatic breast cancer \cite{RN2,RN56}. The pipeline processes high-resolution fluorescence images acquired via the ZEISS$\mathrm{{}^{TM}}$ Axio automated scanning system. High-resolution imaging is performed and an automated algorithm scans the data to identify rare candidates for CTCs among millions of white blood cells\cite{RN2,RN56}. The BRIA machine learning framework filters out non-CTCs and artifacts, reducing the number of candidates for pathologist review\cite{RN2,RN56}. A multiscale feature enhancement algorithm helps identify nuclei while cell segmentation occurs across various fluorescence channels (CK, DAPI, CD45/CD31)\cite{RN2,RN56}. The extracted morphological and molecular characteristics serve as input for a machine learning algorithm trained to identify presumptive CTCs for the validation by experts\cite{RN2,RN56}. Unlike medical-image segmentation approaches that learn directly from image pixels or intermediate image representations, such as the multi-center breast-ultrasound framework of Ye et al.\cite{Ye2026BreastUltrasound}, and the chromosome segmentation/classification methods reviewed by Devaraj et al.\cite{Devaraj2024Chromosome}, our study uses fluorescence digital pathology primarily as a feature-extraction modality. Specifically, CTC images are processed to obtain quantitative morphology, fluorescence-intensity, and texture descriptors, after which QCML operates on these extracted features to predict chromosomal instability/LST rather than performing image segmentation or classification directly on the raw images.

The genomic profiling of CTC was performed as previously published \cite{RN2}. A maximum of 8 CTCs per patient are prioritized by a board-certified pathologist for genomic profiling\cite{RN2}. The selected cells undergo lysis and DNA extraction, followed by amplification of the whole genome and library preparation using the SEQPLEX-I kit\cite{RN2}. Low-pass genome sequencing is used to evaluate chromosomal instability by quantification of LST. A computational pipeline was used to evaluate copy number variations from CTC sequencing data, following principles similar to standard whole-genome sequencing workflows~\cite{RN17}. Sequencing reads generated on the Illumina platform were mapped to the hg38 human reference genome~\cite{dataset_grch38}, and read counts were aggregated in 1-Mb intervals across the genome. Quality control metrics were calculated to exclude samples with low sequencing depth, poor alignment quality, or excessive coverage variability~\cite{RN2}. Only high-quality samples were retained for analysis. To normalize genomic coverage, bin-level read depth was scaled relative to the mean autosomal signal, allowing for correction of chromosome-wide copy number variation\cite{RN2}. 

A total of 112 morphology features were extracted as previously published \cite{RN56}. Eight morphological characteristics were extracted from each of the nuclear and cell masks: 1) size, 2) roundness, 3) elongation, and 4) the first Hu moment \cite{RN57}, to measure a more subtle shape variability.  We next computed 44 intensity features from nuclear and cell masks across the three DAPI, CK and CD45/CD31 channels: 1) MFI, 2) lower, 3) median, and 4) upper quartiles, 5) interquartile range, as well as co-Localizations between channels. 60 texture features were extracted to characterize image patterns. Gabor filters were extracted for localized frequency and orientation information in images and to detect irregularities in repeating textures with a fractal feature approximation. Gabor filters were applied with 16 distinct parameter features combinations, comprising four orientations of the filter ($\theta$ = 0$\mathrm{{}^\circ}$, 45$\mathrm{{}^\circ}$, 90$\mathrm{{}^\circ}$, and 135$\mathrm{{}^\circ}$), two wavelengths (spatial frequency) ($\lambda$ = 0.1 and 0.4), and two standard deviations as Gaussian width ($\sigma$ = 1 and 3), selected based on cell size. For each filtered image, the mean and standard deviation are computed, capturing orientation- and scale-specific frequency content. Another set of features is computed using Laws' texture energy measures \cite{RN61}. This involves generating ordered multiplications of one-dimensional filters---L5 (Level), E5 (Edge), S5 (Spot), and R5 (Ripple)---to detect various spatial patterns, with corresponding statistical descriptors calculated from the filtered outputs. The remaining six features are derived using the Local Binary Pattern (LBP) method, which encodes local texture variations such as edges, corners, and uniform regions. For each image channel, an LBP-transformed image is generated, and inter-channel relationships are quantified using correlation and normalized mutual information, resulting in six final texture features\cite{RN56}.

All patient data were analyzed retrospectively and completely anonymized. All procedures conducted in studies involving samples from human participants adhered to the ethical standards set by the institutional research committee of Epic Sciences, which obtained informed consent from all participants. All experimental protocols involving human samples for CTC analysis were reviewed and approved by the Epic Sciences institutional research committee.

\subsection{Cross-Validation}
\label{sec:cross-val}
For the training and evaluation of all models (QCML and classical ML), we implemented five-fold patient-grouped cross-validation repeated using five distinct random seeds. Patient grouping ensured that all cells from a given patient were assigned exclusively to either the training or validation fold, thereby preventing patient-level leakage between the folds. Identical patient-grouped splits were applied across all model types to enable direct comparisons. Hyperparameters were selected before the repeated evaluation using a subset of 166 CTCs from 51 patients. For each candidate hyperparameter configuration, performance was averaged across five patient-grouped cross-validation folds, and the configuration yielding the highest mean validation performance was selected. Using these fixed hyperparameters, we then performed the complete five-fold, five-repetition patient-grouped cross-validation procedure on the full dataset of 227 CTCs from 73 patients and reported the mean fold-training and fold-validation performance. The same hyperparameter-selection and evaluation procedures were applied consistently across all model types.

\subsection{Quantum Cognition Machine Learning (QCML)}
 \label{sec:qcml}

QCML \cite{CandeloriEtAl, MusaelianEtAl, SamsonEtAl, BlackrockBondSimilarity} is a recently introduced machine learning approach grounded in the principles of quantum cognition (for an overview of quantum cognition, refer to \cite{Porthos_Busemayer_2022}). QCML models represent data observations as quantum states in complex Hilbert space. Recall that in quantum mechanics, a {\em state} is a unit-norm vector in a Hilbert space, defined up to an overall phase. We use the bra-ket notation, representing states by kets such as $\ket{\psi}$. The inner product between two states $\ket{\psi_1}$ and $\ket{\psi_2}$ is denoted by the bra-ket $\braket{\psi_1}{\psi_2}$. A measurement of a quantum observable, represented by a Hermitian operator $M$, in the state $\ket{\psi}$ yields an eigenvalue $m_i$ of $M$ with probability given by the squared magnitude of the overlap with the corresponding eigenstate $\ket{m_i}$: $|\braket{m_i}{\psi}|^2$. The expression $\expval{M}{\psi}$ gives the expected value of the  random variable associated with measuring $M$ in the state $\ket{\psi}$ \cite[I.2.2]{nielsen00}.

In QCML, for each vector $\mathbf{x}_t \in \mathbb{R}^K$ belonging to a data set consisting of $t=1, \ldots, T$ observations, we define an error Hamiltonian as
\begin{equation}
    \label{eq:square_hamiltonian}
    H(\mathbf{x}_{t}) = \frac{1}{2}\sum_k (A_k - \mathbf{x}_{t,k} \cdot I)^2.
\end{equation}
The operators ${A_k} \in \mathbb{C}^{N \times N}$ are a fixed set of quantum observables for $k=1, \ldots, K$, where each observable is represented by a Hermitian operator on an $N$-dimensional Hilbert space. The Hermitian operators are parametrized based on diagonal and off-diagonal entries. For Hermitian operator $A$, the independent components are initialized via a standard normal distribution $\mathcal{N}(0, 1)$:
\[
A_{ii} \sim \mathcal{N}(0, 1) \quad \text{and} \quad A_{ij} = A_{ji}^* = X_{ij} + i Y_{ij} \quad \text{for } i < j, \quad \text{where } X_{ij}, Y_{ij} \sim \mathcal{N}(0, 1)
\]
In Equation \eqref{eq:square_hamiltonian}, $I$ denotes the $N\times N$ identity matrix. Each of these $K$ quantum observables can be viewed as a `quantization' of a corresponding feature of the original $K$-dimensional data set. The vector $\mathbf{x}_t$ then can be mapped to a quantum state $\ket{\psi_t}$ by finding the ground state (i.e., the eigenstate associated with the lowest eigenvalue) of the error Hamiltonian \eqref{eq:square_hamiltonian}. This results in a representation of data into quantum states (i.e., normalized vectors in a complex Hilbert space). Conversely, for an arbitrary quantum state $\ket{\psi}$, its `position' can be defined as the $K$-dimensional real vector
\[
\mathbf{x}(\psi) = \big( \expval{A_k}{\psi} \big)_{k=1}^K \in \mathbb{R}^K.
\]
In quantum theory, this vector represents the expected outcomes of measuring the observables $A_k$ in the quantum state $\ket{\psi}$. As a result, with a set of quantum observables $\{A_k\}$, we can convert data into quantum states by finding the ground state $\ket{\psi_t}$ for each data point $\mathbf{x}_t$, and we can also extract information from any quantum state $\ket{\psi}$ by calculating its position $\mathbf{x}(\psi)$.

In an unsupervised setting, training a QCML model involves iterative updates to the observables $\{A_k\}$ so that the ground states $\ket{\psi_t}$ `cohere' to the data, that is, the distance between $\mathbf{x}_t$ and its position $\mathbf{x}(\psi_t)$ is minimized, as well as the variance of the measurement. During optimization, we can use different forms of a loss function: the distance, the total energy of the error Hamiltonian, or a combined loss function, as discussed in detail in \cite{CandeloriEtAl}.

In the supervised setting, which is the main focus of this article, the training process differs from unsupervised case \cite{BlackrockBondSimilarity}. The target variable $y \in \mathbb{R}$ is assigned a $N$-dimensional quantum `forecast' observable $B$. Given a data point $\mathbf{x}_t$ the corresponding forecast, measured in quantum state $\psi_t$, is given by
\[
\widehat{y}_t = \expval{B}{\psi_t}.
\]
During the training process, the quantum observables $\{A_k\}$ and $B$ are updated at each iteration to minimize a loss function $\mathcal{L}(\widehat{y}_t, y_t)$. The loss function can take various forms such as mean absolute error, mean squared error, cross-entropy, etc. Note that in case of mean absolute error, the non-differentiability of the loss function does not add further complexity to the algorithm, since the mapping from $\mathbf{x}_t$ to its ground state $\psi_t$ is already non-differentiable, the singular points corresponding to the locus of degeneracy of the error Hamiltonian \eqref{eq:square_hamiltonian}. This framework can be easily adapted to handle multiple target variables by introducing separate quantum `forecast' observables for each target. Below is the summary of the training algorithm:

\begin{algorithm}
    \caption*{\bf{QCML univariate regression model training}}
    \label{alg:training}
    \begin{algorithmic}
    \State \textbullet\ Randomly initialize feature operators $\{A_k\}$ and target operator $B$.
    \State \textbullet\ Iterate over training data and operators until desired convergence:
    \begin{algorithmic} \item
        \begin{algorithmic}[1]
        \State Generate error Hamiltonian $H(\mathbf{x}_{t})$
        \State Holding $A_k$ constant, find the ground state $\ket{\psi_t}$ of $H(\mathbf{x}_{t})$
        \State Generate the forecast $\widehat{y}_t = \expval{B}{\psi_t}$
        \State Calculate gradients of the loss function $\mathcal{L}(\widehat{y}_t, y_t)$ w.r.t $A_k$ and $B$
        \State Update $A_k$ and $B$ via gradient descent
        \end{algorithmic}
    \end{algorithmic}
    \end{algorithmic}
\end{algorithm}

The implementation details of these steps vary based on how the operators $A_k$ and $B$ are parameterized, and there are multiple options for loss functions and optimization methods. The Hilbert space dimension $N$ is a hyperparameter that can be tuned through cross-validation. While larger $N$ values generally reduce the loss, they may cause overfitting and poor generalization, whereas smaller dimensions typically result in higher bias but lower variance. \cite{CandeloriEtAl}. For practical purposes, it's also best to keep $N$ small to maintain computational efficiency.

To this end, the main goal is to have a model which produces binary forecast (classification) for a cell being LST positive (LST+) corresponding to LST parameter LST>12, where the cutoff of 12 is based on previously published analytical validation data of the metastatic breast cancer platform \cite{RN2}. We also want the model to 1) produce real-valued LST forecasts, and 2) have the ability to control the balance between specificity and sensitivity. To achieve this, we build a QCML-based regression model and design a mixed-loss function that incorporates both L1 and cross-entropy components, effectively capturing both regression and classification errors. Additionally, the cross-entropy component allows for a varying weight on the positive class to achieve the desired specificity/sensitivity balance. We apply a weight of $w_p=0.5$ to the positive class to prioritize specificity. Below is the outline on generating the forecasts and probabilities:

\begin{enumerate}[label=\arabic*)]
    \item Generate regression forecast: $\widehat{y}_t = \expval{B}{\psi_t}$;
    \item Form true labels for classification: ${{y}_t^p} = \mathbf{1}_{{y}_t > \theta_{LST}}$, where $\theta_{LST}=12$ is our LST threshold;
    \item Form probability forecast for classification: $\widehat{{y}_t^p} = \sigma\Big((\widehat{y}_t - \theta_{LST})s_\sigma\Big)$, where $\sigma(x) = \frac{1}{(1+e^{-x})}$ is the sigmoid function and $s_\sigma$ is a learnable scale parameter.
\end{enumerate}

Then the mixed-loss function takes the following form:

\begin{align}
\mathcal{L}_{\text{Total}} &= 
\frac{\mathcal{L}_{\text{L1}}}{stopgrad(\mathcal{L}_{\text{L1}})} + 
\frac{\mathcal{L}_{\text{CE}}}{stopgrad(\mathcal{L}_{\text{CE}})}, \label{eq:mixed_loss} \\
\text{where} \quad 
\mathcal{L}_{\text{L1}} &= \frac{1}{T} \sum_t |\widehat{y}_t - y_t|, \nonumber \\
\mathcal{L}_{\text{CE}} &= -\frac{1}{T} \sum_t \left[ w_p y_t^p \log(\widehat{y}_t^p) + (1 - y_t^p) \log(1 - \widehat{y}_t^p) \right]. \nonumber
\end{align}

Here, the stop-gradient operation uses the loss values $\mathcal{L}_{\text{L1}}$ and $\mathcal{L}_{\text{CE}}$ in denominator but detaches their gradients within a gradient-descent-based training framework (PyTorch in our case). Essentially, the stop-gradient operation treats denominator values as constant scaling factors during backpropagation, allowing to normalize each loss component to a unit while preserving their gradients. Without the stop-gradient, the numerator and denominator would cancel during differentiation, producing zero gradients. For $\mathcal{L}_{\text{Total}}$, adjusting each loss component by the corresponding stop-gradient loss forces each component's loss to be equal to 1. However, the gradient $\nabla \mathcal{L}_{\text{Total}}$ will not necessarily be 0, allowing the model to learn while maintaining consistent weighting between the loss components.

\subsection{QCML Positive Operator-Valued Measure}
 \label{sec:qcml_povm}

QCML Positive Operator-Valued Measure (POVM) extends QCML to predict probability density functions for targets instead of single point estimates. This extension provides the ability to forecast full probability distributions, which enables the estimation of confidence intervals and offers a more detailed understanding of the predictions. This approach is particularly well-suited for our LST forecasting task, as it allows us to produce both continuous-valued predictions (e.g., expected mean or median LST) and probabilistic forecasts (e.g., the likelihood that LST exceeds a specified threshold).

Generalized measurements in quantum mechanics are described by a set of operators known as a Positive Operator-Valued Measure (POVM)~\cite{nielsen00}. A POVM is a collection $\{\hat{F}_k\}$ of positive semi-definite operators acting on the Hilbert space, such that $\sum_k \hat{F}_k = \hat{I}$. Each operator $\hat{F}_k$  corresponds to a possible measurement outcome. The connection between the quantum state and the measurement outcomes is provided by Born's rule. For a state $ \ket{\psi}$, the probability of observing the outcome associated with the POVM element $k$ is given by $p_k = \bra{\psi}\hat{F}_k\ket{\psi}$. 

QCML POVM allows us to forecast the probability density function $p(y)$ of a continuous target variable $y$ instead of point estimates. Without loss of generality, we will assume that $y \in [-1, 1]$. Suppose that we want to generate a probability density function of a continuous variable $y$ conditional on a quantum state $\ket{\psi}$.  We introduce a function mapping $y$ into output operators $\hat{Y}(y)$, generally non-Hermitian, such that 
    \begin{align}\label{povm}
        \int_{-1}^1 \hat{Y}^{\dagger}(y)\hat{Y}(y)dy = \hat{I}.
    \end{align}
The set of operators $\hat{F}(y) = \hat{Y}^\dagger(y)\hat{Y}(y)$, indexed by the continuous parameter $y$, forms a POVM. By construction, the probability density function $p(y)$ is given by
\begin{align}\label{pdf}
    p(y) = \bra{\psi}\hat{Y}^{\dagger}(y)\hat{Y}(y)\ket{\psi}.
\end{align}
 The POVM elements $\hat{Y}(y)$ can be parametrized in a variety of ways. Here, we suggest parametrization in terms of a finite number of Legendre polynomials \cite{weber2005mathematical} $L_k(y)$:
\begin{align*}
    \hat{Y}(y) = \sum_{n=0}^{K-1}\hat{A}_n L_n(y)\sqrt{\frac{2n+1}{2}},
\end{align*}
where, $K$ is the truncation parameter and $\hat{A}_k$ are generally non-Hermitian matrices to be learned. 
Then Equation (\ref{pdf}) becomes:
\begin{align*}
    p(y) = \sum_{n,m}\bra{\psi}\hat{A}_n^{\dagger}\hat{A}_{m}\ket{\psi}L_n(y)L_{m}(y)\sqrt{\frac{2n+1}{2}}\sqrt{\frac{2m+1}{2}}.
\end{align*}
Given the orthogonality of Legendre polynomials:
\begin{align}\label{orthonormality}
    \int_{-1}^1 L_n(z)L_m(z)dz = \frac{2}{2n+1} \delta_{nm},
\end{align} 
it follows from Equation (\ref{povm}) that:
\begin{align*}
    \sum_{n=0}^{K-1}\hat{A}_n^{\dagger}\hat{A}_n = \hat{I}.
\end{align*}
Therefore, the matrices $\hat{F}_n=\hat{A}_n^\dagger\hat{A}_n$ form a POVM.

For a target variable $y$ on arbitrary support $[a, b]$, we consider a PDF $g(y)$ as an initial guess. We use this distribution to do a variable transformation into $z \in [-1, 1]$:
\begin{align*}
    z = G(y) = 2\int_a^y g(t)dt - 1.
\end{align*}
Now we construct the PDF to be learned as:
\begin{align*}
    p(y) = 2\sum_{nm}\sqrt{\frac{2n+1}{2}}\sqrt{\frac{2m+1}{2}}\bra{\psi}\hat{A}_n^{\dagger}\hat{A}_m\ket{\psi}L_n(G(y))L_m(G(y))g(y).
\end{align*}
Using (\ref{orthonormality}) and making a variable transformation $z = G(y)$ and $dz = 2g(y)dy$ we can confirm that
\begin{align*}
    \int_a^b p(y)dy = \sum_n \bra{\psi}\hat{A}_n^{\dagger}\hat{A}_n\ket{\psi} = 1.
\end{align*}
As a special case, if  
$$
\langle\psi|\hat A_n^\dagger\hat A_m|\psi\rangle
=\delta_{n0}\delta_{m0},
\qquad n,m=0,\ldots,K-1,
$$
then only the $n=m=0$ term contributes and $p(y)=g(y)$.

\subsection{Code Availability}
\label{sec:code_availability}
The community version of the QCML package is available on PyPI at \url{https://pypi.org/project/hone-io/} and can be installed through \texttt{pip install hone-io==2.3.0}. The community version of the package has some limitations on dataset size (max input features: 100, max samples: 1000) and Hilbert space dimensions (max 8). Our study uses 112 input features and to repeat the experiments the enterprise version of the package is required. The enterprise version (\url{https://pypi.org/project/hone-io-enterprise/}) does not have any size limitations. To access the enterprise version  please email us at \href{mailto:support@qognitive.io}{support@qognitive.io}. The enterprise version can be installed through \texttt{pip install hone-io-enterprise==2.3.0}.  The package documentation can be found at \url{https://qognitive-qcog-core.readthedocs-hosted.com/en/latest/}.

\section{Results}
\label{sec:results}

\subsection{CTC Dataset Description}
\begin{figure}
    \centering
    \includegraphics[width=1\linewidth]{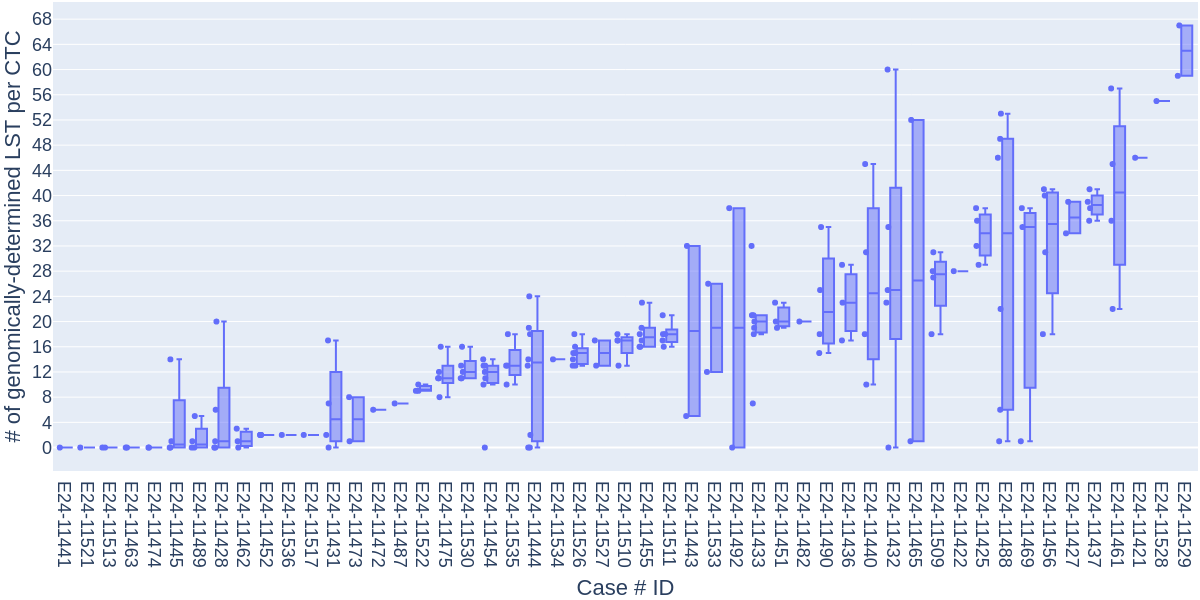}
    \caption{\textbf{Distribution of genomically determined LST values across all cases.} Scatter plots show the genomically determined LST ground truth values which are shown per CTC, sub-grouped by case ID number and ranked from left to right by increasing LST values. We show the visual representation of the heterogeneity in the distribution of LST values across each case ID.}
    \label{fig:LST_distribution_by_case}
\end{figure}

\begin{figure}
    \centering
    \begin{subfigure}{.89\textwidth}
        \centering
        \includegraphics[width=\linewidth]{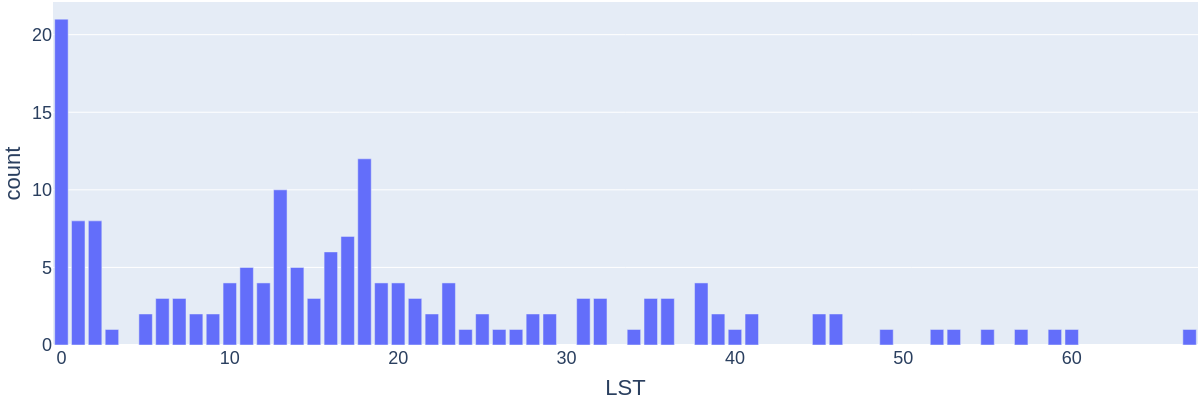}
        \label{fig:LST_value_count_actual}
        \captionsetup{font={small}}
        \caption{Genomically determined LST.}
    \end{subfigure}
    \begin{subfigure}{.89\textwidth}
        \centering
        \includegraphics[trim=0 0 0 0, clip, width=\linewidth]{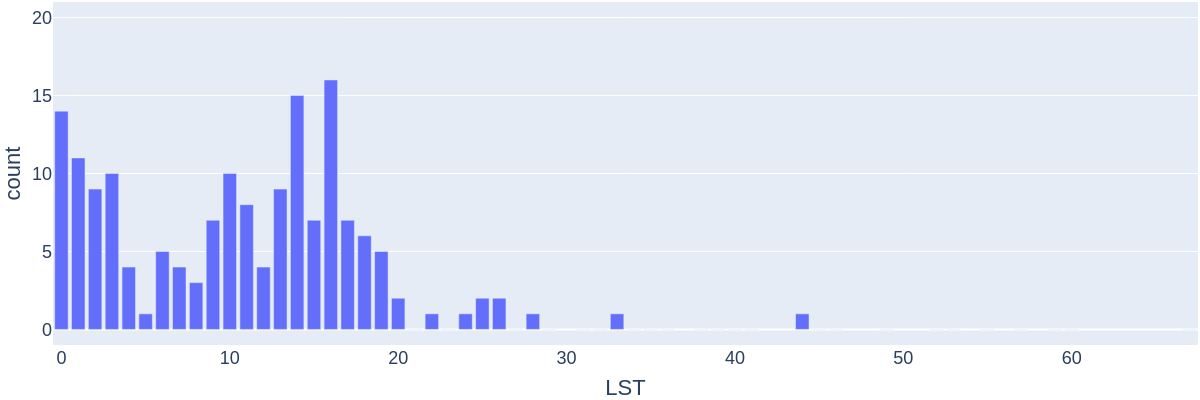}
        \label{fig:LST_value_count_forecast}
        \captionsetup{font={small}}
        \caption{Morphology-predicted LST.}
    \end{subfigure}
    \caption{\textbf{Count of CTCs with specific LST values.} The bar plot comparison illustrates the number of CTCs ranked from left to right by increasing LST values. The bar plot at the top displays the LST values ground truth, which is the genomically determined LST per CTC, while the bottom plot shows the morphology-predicted LST values computed by QCML. The count of CTCs for the ground truth genomically determined LST follows the same bimodal trend as that of the morphology-predicted LST.}
    \label{fig:LST_value_count}
\end{figure}

A total of 227 available cells were identified across all patients. These cells were 1) selected as candidate CTC by a trained pathologist, 2) sequenced, and 3) met the genomic quality control (QC) metrics established by our pipeline \cite{RN2}. On average, each patient had 3.1 sequenced CTCs (with a standard deviation of 1.8; the maximum was eight, and the minimum was 1). This data established the ground truth for measuring chromosomal instability based on the number of LST. We divided the overall cohort into a development subset, which included 51 patients and 166 CTCs. \autoref{fig:LST_distribution_by_case} shows the heterogeneity of LST values across each case in the development subset. \autoref{fig:LST_value_count}a shows the genomically determined LST values distribution for all cases, organized by case ID number and ranked in ascending order based on LST values. We identify that 73\% (37 out of 51) of the cases in the development subset had at least one LST+ CTC. Following that, digital pathology features, including cell morphology and fluorescence intensity levels were extracted from each CTC image. As detailed in \nameref{sec:methods} Section, and previously published \cite{RN56} 112 morphology features were extracted: 8 from nuclear and cell masks (size, roundness, elongation, and the first Hu moment) and 44 intensity features from DAPI, CK, and CD45/CD31 channels (MFI, lower, median, upper quartiles, interquartile range, and co-localizations). Additionally, 60 texture features were obtained using Gabor filters, applied with 16 parameters to capture frequency content, orientation, and irregularities with fractal approximations. Laws' texture energy measures were used for detecting spatial patterns, and six features were derived from the Local Binary Pattern method to analyze local texture variations across image channels. 

\begin{figure}[H]
    \centering
    \includegraphics[width=\linewidth]{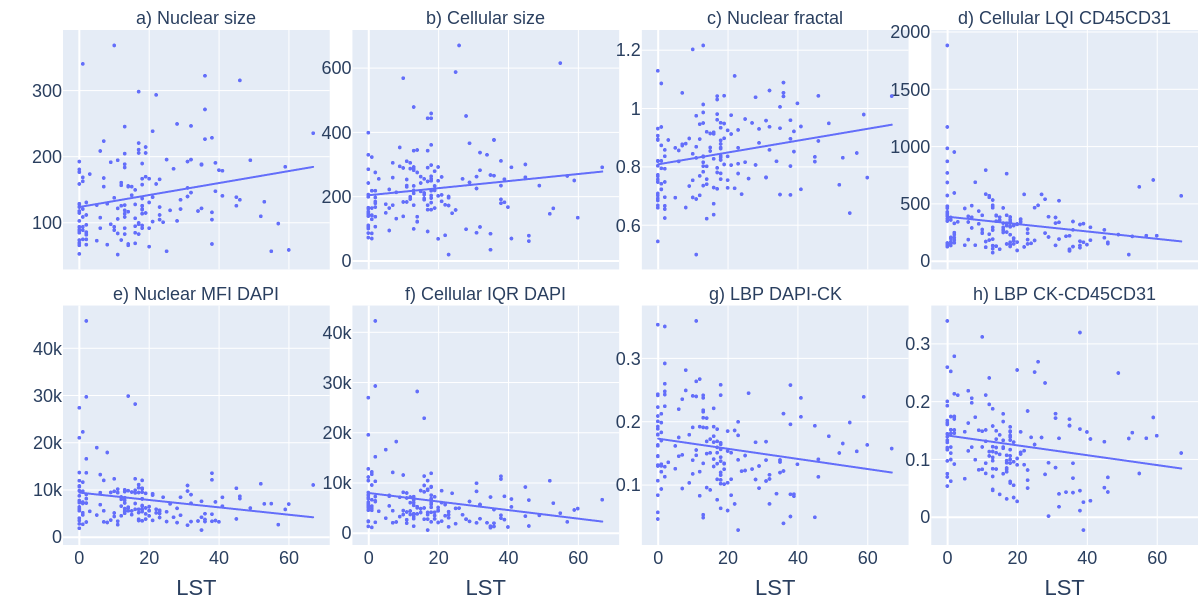}
    \caption{\textbf{Linear correlation of cellular and nuclear morphology and protein intensity features with the extent of LST as determined by genome sequencing. }In the scatter plots each dot is an individual CTC and graphs show the best-fit line between the extent of LST as a measure of chromosomal instability and a) nuclear size, b) cellular size, c) nuclear fractal, d) cellular LQI CD45/CD31, e) nuclear MFI DAPI, f) cellular IQR DAPI, g) LBP DAPI-CK, h) LBP CK-CD45/CD31.}
    \label{fig:feature_corr_to_LST}
\end{figure}

To understand the linear relationships between metastatic breast cancer CTCs digital pathology features and their ground truth genomically determined LST, we calculated Pearson's correlations between them. We found that CTCs with a higher degree of chromosomal instability, represented by higher LST values, were significantly correlated with a larger nuclear $(r=0.22, P=0.032)$ and cellular $(r=0.15, P=0.04)$ morphology size, which is a measure of the overall pixel area of a segmented nucleus and cell (\autoref{fig:feature_corr_to_LST}a and \ref{fig:feature_corr_to_LST}b). It was also substantially correlated with nuclear fractal features which measure shape complexity and heterogeneity $(r=0.24, P=0.001)$ (\autoref{fig:feature_corr_to_LST}c). These results are consistent with nuclear enlargements, spatial disorganization and pleomorphism which may be due to polyploidy, and multinucleation which is expected to occur in genomically unstable cells \cite{RN81,RN1,RN83}. Several CD45/CD31 intensity values were correlated with lower LST with the most significant being cellular low quartile range LQI $(r=-0.21, P=0.005)$ (\autoref{fig:feature_corr_to_LST}d). CD45/CD31 is a negative CTC marker that is typically down-regulated in cancer cells of epithelial origin from solid tumors. LST values trended with a lower expression of several DAPI intensities which were most correlated as expressed by mean fluorescent intensity (MFI) in the nuclear mask $(r=-0.21, P=0.0072)$ (\autoref{fig:feature_corr_to_LST}e) and the interquartile range IQI in the cellular mask $(r=-0.23, P=0.002)$ (\autoref{fig:feature_corr_to_LST}f). DAPI binds strongly to A-T rich regions of double stranded DNA~\cite{RN69}. However, the inverse correlation of DAPI nuclear and cellular intensities result may be explained by the fact that genomically unstable cells are expected
to show unpredictable intensity patterns due to chromatin remodeling, micronuclei, or fragmentation \cite{RN63,RN64,RN69}. Also, DAPI intensity can be affected by cell cycle phases as G2/M cells are expected to have more DNA content than G1 phase which are typically altered during chromosomal instability \cite{RN63,RN64,RN69}. Cross-channel Local Binary Pattern (LBP) for CK-DAPI and for CK-CD45/CD31 channel pairs, which is a measure of cross channel correlation and similarity across local binary pattern for channel pairs within a segmented cell \cite{RN62,RN56}, were found to be significantly inversely correlated with the extent of LST numbers $(r=-0.19, P=0.01; r=-0.21, P=0.008)$ (\autoref{fig:feature_corr_to_LST}g and \ref{fig:feature_corr_to_LST}h). The cross-channel LBP is a measure of colocalization between proteins of a CTC calculated by comparing pixels intensity in the same position for each of the two channels. In doing so, one can capture subtle spatial relationships and structural changes such as nuclear deformities and cytoskeletal remodeling \cite{RN62,RN56}. Therefore, a lower texture cross-channel LBP suggests spatial discordance between nuclear and cytoplasmic structure, which is expected during instability-driven morphological shifts.  

\subsection{Regression Models}
We train a QCML model along with classical machine learning models (see \autoref{tab:QCML_performance}) following the cross-validation procedure described in \Cref{sec:cross-val}. The problem is set up as a regression task in which we produce a real-valued LST forecast and then apply a cutoff of LST > 12 to produce a binary forecast of CTCs being LST+. As described in Equation (\ref{eq:mixed_loss}) in \Cref{sec:qcml}, our QCML model employs a mixed loss function and allows for varying weights on the positive class, enabling a balance between specificity and sensitivity. The incidence of false positive cases is particularly detrimental in clinical settings and needs to be minimized as much as possible. It may erroneously signal a lack of response from current treatment and lead to an unnecessary change in an otherwise effective line of therapy. For this reason, a weight of 0.5 was applied to the positive class to prioritize specificity and reduce the occurrence of false positive cases. To test the hypothesis that QCML can improve prediction performance and reproducibility across independent datasets, we compared it with 12 machine-learning models: TabPFN, CatBoost, SVM with a Gaussian kernel, Elastic Net, linear SVM, XGBoost, multilayer perceptron (MLP), AdaBoost, nearest neighbors with 5 and 32 neighbors, Random Forest, and linear regression \cite{RN27} (\autoref{tab:QCML_performance}).

\begin{table}[H]
\centering
\scriptsize
\renewcommand{\arraystretch}{1.2}
\begin{tabularx}{\textwidth}{|l|Y|Y|Y|Y|Y|Y|}
\hline
\textbf{Models} & \multicolumn{3}{c|}{\textbf{Fold training}} & \multicolumn{3}{c|}{\textbf{Fold validation}} \\
\cline{2-7}
 & \textbf{Sensitivity} & \textbf{Specificity} & \textbf{Balanced Accuracy} & \textbf{Sensitivity} & \textbf{Specificity} & \textbf{Balanced Accuracy} \\
\hline
\textbf{QCML} & 91\% {\tiny$\pm$ 3\%} & 67\% {\tiny$\pm$ 6\%} & 79\% {\tiny$\pm$ 3\%} & 81\% {\tiny$\pm$ 10\%} & \textbf{53\%} {\tiny$\pm$ 16\%} & \textbf{67\%} {\tiny$\pm$ 9\%} \\
\hline
\textbf{TabPFN}\cite{hollmann2025accurate} & \textbf{100\%} {\tiny$\pm$ 0\%} & 99\% {\tiny$\pm$ 2\%} & 99\% {\tiny$\pm$ 1\%} & 90\% {\tiny$\pm$ 8\%} & 41\% {\tiny$\pm$ 16\%} & 66\% {\tiny$\pm$ 8\%} \\
\hline
\textbf{CatBoost}\cite{Prokhorenkova2018CatBoost} & \textbf{100\%} {\tiny$\pm$ 0\%} & \textbf{100\%} {\tiny$\pm$ 1\%} & \textbf{100\%} {\tiny$\pm$ 1\%} & 93\% {\tiny$\pm$ 7\%} & 36\% {\tiny$\pm$ 17\%} & 64\% {\tiny$\pm$ 8\%} \\
\hline
\textbf{SVM Gaussian Kernel} & 98\% {\tiny$\pm$ 1\%} & 73\% {\tiny$\pm$ 5\%} & 85\% {\tiny$\pm$ 2\%} & 90\% {\tiny$\pm$ 7\%} & 39\% {\tiny$\pm$ 16\%} & 64\% {\tiny$\pm$ 8\%} \\
\hline
\textbf{Elastic Net} & 95\% {\tiny$\pm$ 2\%} & 52\% {\tiny$\pm$ 8\%} & 73\% {\tiny$\pm$ 4\%} & 86\% {\tiny$\pm$ 8\%} & 40\% {\tiny$\pm$ 16\%} & 63\% {\tiny$\pm$ 8\%} \\
\hline
\textbf{Linear SVM} & 93\% {\tiny$\pm$ 2\%} & 58\% {\tiny$\pm$ 6\%} & 75\% {\tiny$\pm$ 3\%} & 84\% {\tiny$\pm$ 7\%} & 42\% {\tiny$\pm$ 14\%} & 63\% {\tiny$\pm$ 8\%} \\
\hline
\textbf{XGBoost} & \textbf{100\%} {\tiny$\pm$ 0\%} & 99\% {\tiny$\pm$ 1\%} & \textbf{100\%} {\tiny$\pm$ 1\%} & 82\% {\tiny$\pm$ 11\%} & 41\% {\tiny$\pm$ 15\%} & 61\% {\tiny$\pm$ 8\%} \\
\hline
\textbf{MLP} & 97\% {\tiny$\pm$ 2\%} & 91\% {\tiny$\pm$ 7\%} & 94\% {\tiny$\pm$ 3\%} & 77\% {\tiny$\pm$ 8\%} & 52\% {\tiny$\pm$ 15\%} & 65\% {\tiny$\pm$ 7\%} \\
\hline
\textbf{AdaBoost} & 97\% {\tiny$\pm$ 1\%} & 70\% {\tiny$\pm$ 4\%} & 84\% {\tiny$\pm$ 2\%} & 91\% {\tiny$\pm$ 8\%} & 41\% {\tiny$\pm$ 17\%} & 66\% {\tiny$\pm$ 8\%} \\
\hline
\textbf{Nearest Neighbors 5} & 95\% {\tiny$\pm$ 2\%} & 46\% {\tiny$\pm$ 6\%} & 71\% {\tiny$\pm$ 3\%} & 89\% {\tiny$\pm$ 7\%} & 36\% {\tiny$\pm$ 16\%} & 63\% {\tiny$\pm$ 8\%} \\
\hline
\textbf{Random Forest} & 98\% {\tiny$\pm$ 1\%} & 69\% {\tiny$\pm$ 4\%} & 84\% {\tiny$\pm$ 2\%} & 91\% {\tiny$\pm$ 9\%} & 35\% {\tiny$\pm$ 16\%} & 63\% {\tiny$\pm$ 8\%} \\
\hline
\textbf{Nearest Neighbors 32} & 98\% {\tiny$\pm$ 1\%} & 19\% {\tiny$\pm$ 10\%} & 59\% {\tiny$\pm$ 5\%} & \textbf{97\%} {\tiny$\pm$ 4\%} & 15\% {\tiny$\pm$ 10\%} & 56\% {\tiny$\pm$ 4\%} \\
\hline
\textbf{Linear Regression} & 94\% {\tiny$\pm$ 2\%} & 76\% {\tiny$\pm$ 6\%} & 85\% {\tiny$\pm$ 3\%} & 69\% {\tiny$\pm$ 10\%} & 43\% {\tiny$\pm$ 13\%} & 56\% {\tiny$\pm$ 8\%} \\
\hline
\end{tabularx}
\caption{\textbf{Fold-training and fold-validation performance of QCML and classical ML models forecasting LST+.} Showing the average and standard deviation of sensitivity, specificity and balanced accuracy across 25 folds (5-fold with 5 repeats).}
\label{tab:QCML_performance}
\end{table}

QCML achieved the highest mean fold-validation specificity (53\%) and balanced accuracy (67\%), while maintaining a sensitivity of 81\%. QCML also showed a smaller difference between fold-training and fold-validation performance than several comparison models, suggesting less overfitting in this dataset. Specifically, QCML achieved fold-training and fold-validation balanced accuracies of 79\% and 67\%, respectively, compared with 94\% and 65\% for the MLP. Both QCML and MLP can learn intermediate representations; in QCML, these representations are encoded by the learned quantum states. In \autoref{fig:LST_value_count}b, we also show the distribution of QCML morphology-predicted LST values across all CTCs, which follows a trend similar to that of the genomically determined LST values shown in \autoref{fig:LST_value_count}a.

\subsection{Classification Models}
Here, as opposed to a regression model, we set up the problem as a classification task where we forecast a binary target of CTCs being LST+. Although this approach does not generate real-valued forecasts, it allows evaluating models with various probability thresholds to target a specific balance between specificity and sensitivity. Additionally, it allows measurement of threshold-independent metrics like AUC-ROC to summarize the performance across all possible classification thresholds. For QCML we use the \nameref{sec:qcml_povm} model to produce probability forecasts, where we model the LST target based on an exponential transformation and use a Legendre polynomial parametrization. \autoref{tab:POVM_performance} shows the performance of QCML POVM in conjunction with classical machine learning models based on a probability threshold of 0.6~\cite{Youden1950, Perkins2006}.

\begin{table}[H]
\centering
\scriptsize
\renewcommand{\arraystretch}{1.2}
\begin{tabularx}{\textwidth}{|l|Y|Y|Y|Y|Y|Y|Y|Y|}
\hline
 & \multicolumn{4}{c|}{\textbf{Fold training}} & \multicolumn{4}{c|}{\textbf{Fold validation}} \\
\cline{2-9}
\textbf{Model} & \textbf{Sensitivity} & \textbf{Specificity} & \textbf{Balanced Accuracy} & \textbf{ROC AUC} & \textbf{Sensitivity} & \textbf{Specificity} & \textbf{Balanced Accuracy} & \textbf{ROC AUC} \\
\hline
\textbf{QCML POVM} & \mbox{92\% {\tiny$\pm$ 3\%}} & \mbox{83\% {\tiny$\pm$ 6\%}} & \mbox{87\% {\tiny$\pm$ 3\%}} & \mbox{0.946} & \mbox{76\% {\tiny$\pm$ 9\%}} & \mbox{58\% {\tiny$\pm$ 18\%}} & \mbox{67\% {\tiny$\pm$ 9\%}} & \mbox{\textbf{0.739}} \\
\hline
\textbf{QCML} & \mbox{81\% {\tiny$\pm$ 5\%}} & \mbox{85\% {\tiny$\pm$ 4\%}} & \mbox{83\% {\tiny$\pm$ 2\%}} & \mbox{0.922} & \mbox{67\% {\tiny$\pm$ 13\%}} & \mbox{62\% {\tiny$\pm$ 18\%}} & \mbox{65\% {\tiny$\pm$ 9\%}} & \mbox{0.710} \\
\hline
\textbf{CatBoost} & \mbox{\textbf{100\%} {\tiny$\pm$ 0\%}} & \mbox{\textbf{100\%} {\tiny$\pm$ 0\%}} & \mbox{\textbf{100\%} {\tiny$\pm$ 0\%}} & \mbox{\textbf{1.000}} & \mbox{\textbf{80\%} {\tiny$\pm$ 10\%}} & \mbox{55\% {\tiny$\pm$ 18\%}} & \mbox{\textbf{68\%} {\tiny$\pm$ 8\%}} & \mbox{0.724} \\
\hline
\textbf{TabPFN} & \mbox{\textbf{100\%} {\tiny$\pm$ 0\%}} & \mbox{\textbf{100\%} {\tiny$\pm$ 0\%}} & \mbox{\textbf{100\%} {\tiny$\pm$ 0\%}} & \mbox{\textbf{1.000}} & \mbox{76\% {\tiny$\pm$ 10\%}} & \mbox{57\% {\tiny$\pm$ 18\%}} & \mbox{66\% {\tiny$\pm$ 6\%}} & \mbox{0.714} \\
\hline
\textbf{XGBoost} & \mbox{\textbf{100\%} {\tiny$\pm$ 0\%}} & \mbox{\textbf{100\%} {\tiny$\pm$ 0\%}} & \mbox{\textbf{100\%} {\tiny$\pm$ 0\%}} & \mbox{\textbf{1.000}} & \mbox{76\% {\tiny$\pm$ 11\%}} & \mbox{55\% {\tiny$\pm$ 18\%}} & \mbox{66\% {\tiny$\pm$ 7\%}} & \mbox{0.715} \\
\hline
\textbf{Random Forest} & \mbox{96\% {\tiny$\pm$ 1\%}} & \mbox{\textbf{100\%} {\tiny$\pm$ 0\%}} & \mbox{98\% {\tiny$\pm$ 1\%}} & \mbox{0.999} & \mbox{65\% {\tiny$\pm$ 16\%}} & \mbox{60\% {\tiny$\pm$ 17\%}} & \mbox{63\% {\tiny$\pm$ 8\%}} & \mbox{0.708} \\
\hline
\textbf{Nearest Neighbors 32} & \mbox{82\% {\tiny$\pm$ 5\%}} & \mbox{59\% {\tiny$\pm$ 8\%}} & \mbox{71\% {\tiny$\pm$ 2\%}} & \mbox{0.788} & \mbox{\textbf{80\%} {\tiny$\pm$ 12\%}} & \mbox{54\% {\tiny$\pm$ 17\%}} & \mbox{67\% {\tiny$\pm$ 9\%}} & \mbox{0.706} \\
\hline
\textbf{Nearest Neighbors 5} & \mbox{71\% {\tiny$\pm$ 5\%}} & \mbox{90\% {\tiny$\pm$ 4\%}} & \mbox{80\% {\tiny$\pm$ 3\%}} & \mbox{0.887} & \mbox{60\% {\tiny$\pm$ 11\%}} & \mbox{63\% {\tiny$\pm$ 14\%}} & \mbox{62\% {\tiny$\pm$ 7\%}} & \mbox{0.693} \\
\hline
\textbf{RBF SVM} & \mbox{80\% {\tiny$\pm$ 6\%}} & \mbox{65\% {\tiny$\pm$ 6\%}} & \mbox{73\% {\tiny$\pm$ 2\%}} & \mbox{0.817} & \mbox{72\% {\tiny$\pm$ 14\%}} & \mbox{60\% {\tiny$\pm$ 18\%}} & \mbox{66\% {\tiny$\pm$ 9\%}} & \mbox{0.714} \\
\hline
\textbf{MLP} & \mbox{\textbf{100\%} {\tiny$\pm$ 0\%}} & \mbox{\textbf{100\%} {\tiny$\pm$ 0\%}} & \mbox{\textbf{100\%} {\tiny$\pm$ 0\%}} & \mbox{\textbf{1.000}} & \mbox{79\% {\tiny$\pm$ 8\%}} & \mbox{54\% {\tiny$\pm$ 15\%}} & \mbox{66\% {\tiny$\pm$ 7\%}} & \mbox{0.672} \\
\hline
\textbf{Linear SVM} & \mbox{85\% {\tiny$\pm$ 4\%}} & \mbox{70\% {\tiny$\pm$ 7\%}} & \mbox{77\% {\tiny$\pm$ 3\%}} & \mbox{0.864} & \mbox{76\% {\tiny$\pm$ 12\%}} & \mbox{54\% {\tiny$\pm$ 17\%}} & \mbox{65\% {\tiny$\pm$ 10\%}} & \mbox{0.694} \\
\hline
\textbf{AdaBoost} & \mbox{37\% {\tiny$\pm$ 7\%}} & \mbox{\textbf{100\%} {\tiny$\pm$ 0\%}} & \mbox{69\% {\tiny$\pm$ 3\%}} & \mbox{0.999} & \mbox{24\% {\tiny$\pm$ 13\%}} & \mbox{\textbf{91\%} {\tiny$\pm$ 10\%}} & \mbox{57\% {\tiny$\pm$ 6\%}} & \mbox{0.686} \\
\hline
\textbf{Logistic Regression} & \mbox{87\% {\tiny$\pm$ 3\%}} & \mbox{86\% {\tiny$\pm$ 4\%}} & \mbox{87\% {\tiny$\pm$ 3\%}} & \mbox{0.945} & \mbox{71\% {\tiny$\pm$ 10\%}} & \mbox{55\% {\tiny$\pm$ 17\%}} & \mbox{63\% {\tiny$\pm$ 7\%}} & \mbox{0.641} \\
\hline
\textbf{Naive Bayes} & \mbox{64\% {\tiny$\pm$ 14\%}} & \mbox{71\% {\tiny$\pm$ 13\%}} & \mbox{67\% {\tiny$\pm$ 3\%}} & \mbox{0.749} & \mbox{57\% {\tiny$\pm$ 17\%}} & \mbox{58\% {\tiny$\pm$ 18\%}} & \mbox{58\% {\tiny$\pm$ 9\%}} & \mbox{0.622} \\
\hline
\end{tabularx}
\caption{\textbf{Fold-training and fold-validation performance of QCML POVM and classical ML classification models forecasting LST+.} Showing the average and standard deviation of sensitivity, specificity and balanced accuracy across 25 folds (5-fold with 5 repeats). Also showing the average ROC AUC score per model.}
\label{tab:POVM_performance}
\end{table}

\begin{figure}[H]
    \centering
    \includegraphics[width=.9\linewidth]{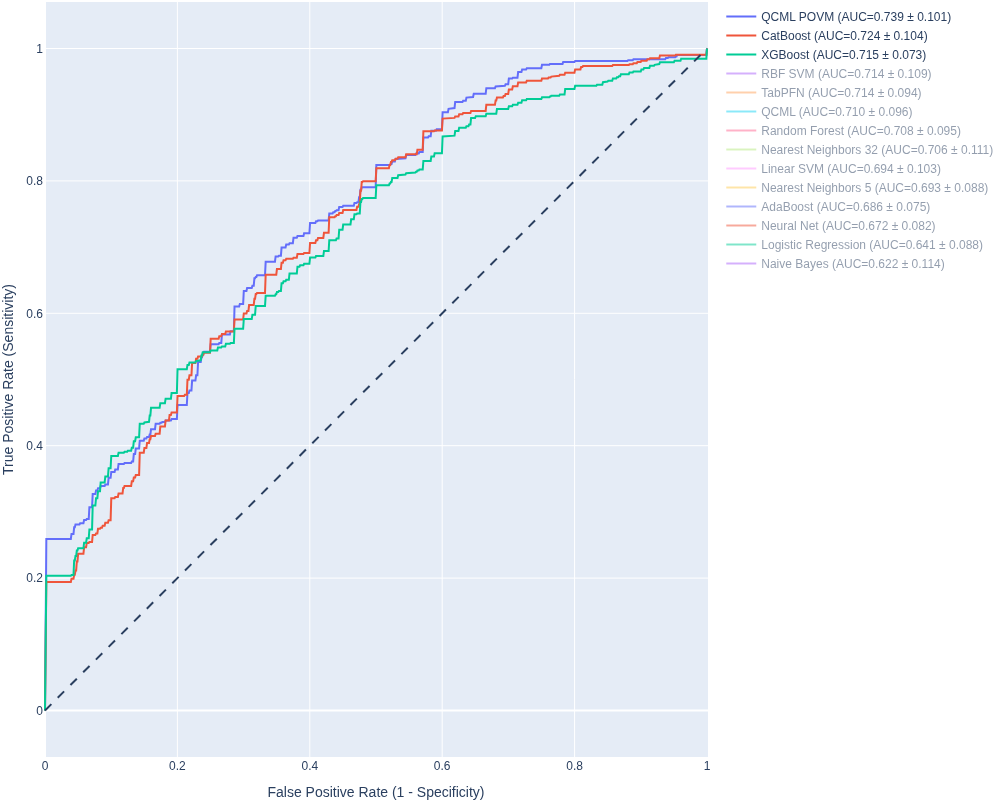}
    \caption{\textbf{ROC curves for the top three performing models}: QCML POVM (blue), CatBoost (red), and XGBoost (green). At $\sim$70\% specificity all three models achieve similar sensitivity. At lower specificity, QCML POVM is outperforming both CatBoost and XGBoost (i.e., it can achieve higher sensitivity for a fixed specificity). At higher specificity, QCML POVM is on par with XGBoost and outperforms CatBoost.}
    \label{fig:roc_auc}
\end{figure}

\begin{figure}[H]
    \centering
    \includegraphics[trim=0 10 0 70, clip, width=.9\linewidth]{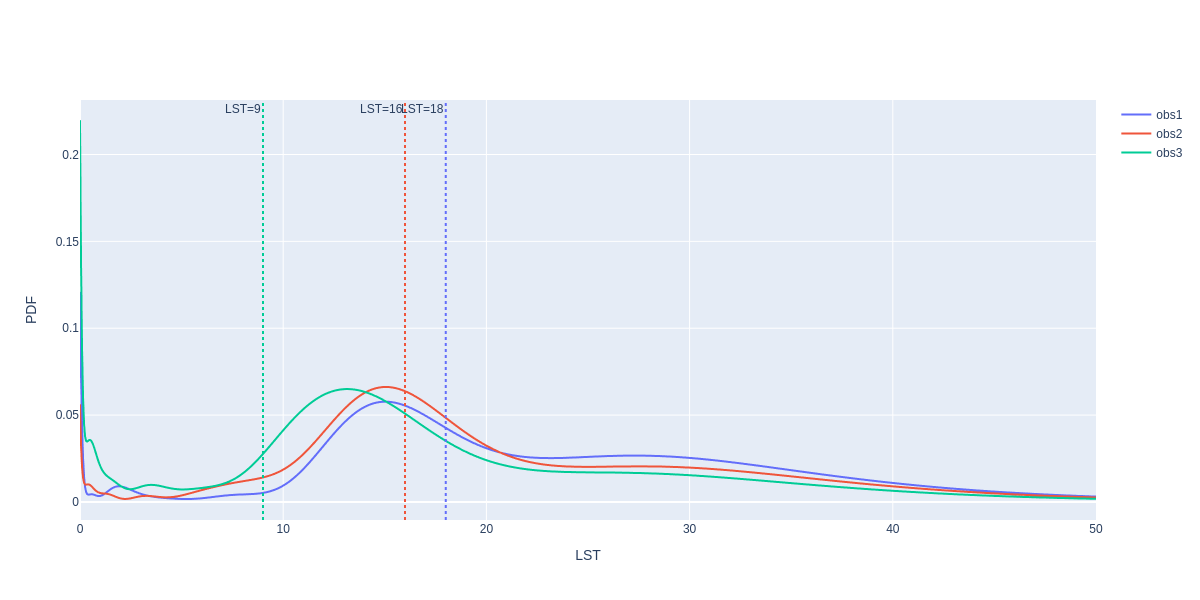}
    \caption{\textbf{QCML POVM produces full probability distribution of LST for all observations.} Showing predicted LST distribution for a sample of 3 CTCs. Dotted vertical lines show the actual LST for each CTC. Having probability distribution allows one to forecast both LST quantiles as well as probabilities for LST being above/below a threshold.}
    \label{fig:POVM_LST_distribution}
\end{figure}

QCML POVM achieved the highest mean fold-validation ROC--AUC (0.739) among the evaluated classification models, as shown in \autoref{tab:POVM_performance} and \autoref{fig:roc_auc}. Additionally, QCML POVM generated full probability distributions of LST values, as shown in \autoref{fig:POVM_LST_distribution}.
\vspace{-.1in}

\subsection{Model specifications and sensitivities}

The main hyperparameters of a QCML model are the Hilbert space dimension $N$, number of epochs and the learning rate. For the main results reported in Tables 1 and 2, we used an ensemble of two QCML models with $N=4$ and $N=8$. The final prediction for each observation was obtained by taking the arithmetic mean of the predictions from the two models. We used 500 epochs, a learning rate of $0.001$, and the Adam optimizer for training. All the results and figures for this article were obtained on a laptop equipped with a 32-core 13th Generation Intel Core i9-13950HX processor with 64 GB of memory, supplemented by an NVIDIA RTX 5000 Ada Generation Laptop graphics processing unit (GPU).

To evaluate sensitivity to the choice of Hilbert-space dimension, we also trained individual models with $N=4$, $8$, $16$, and $32$. \autoref{tab:hs-dims-performance} shows similar performance across these dimensions. The models with $N=16$ and $N=32$ were included only in this sensitivity analysis and were not used in the ensemble.

\begin{table}[H]
\centering
\scriptsize
\renewcommand{\arraystretch}{1.2}
\begin{tabularx}{0.5\textwidth}{|c|Y|}
\hline
\textbf{HS Dims} & \textbf{Balanced Accuracy : Fold validation} \\
\hline
\textbf{4} & 67.3\% \\
\hline
\textbf{8} & 67.6\% \\
\hline
\textbf{16} & 67.3\% \\
\hline
\textbf{32} & 66.7\% \\
\hline
\end{tabularx}
\caption{\textbf{QCML performance across Hilbert Space dimensions.}}
\label{tab:hs-dims-performance}
\end{table}

For clinical deployment, efficiency of the forecasting algorithm can be critical. \autoref{tab:runtime_performance} compares QCML runtime with better performing state-of-the-art models. Although QCML required more training time, it had the shortest inference time in the reported benchmark, supporting its potential suitability for applications requiring rapid inference.

\begin{table}[H]
\centering
\scriptsize
\renewcommand{\arraystretch}{1.2}
\begin{tabularx}{0.55\textwidth}{|l|Y|Y|}
\hline
\multicolumn{3}{|c|}{\textbf{Runtime Comparison}} \\
\hline
\textbf{Model} & \textbf{Train Runtime} & \textbf{Inference Runtime (100,000 cells)} \\
\hline
\textbf{QCML} & 120s & 0.01s \\
\hline
\textbf{CatBoost} & 30s & 0.05s \\
\hline
\textbf{TabPFN} & 26s & 59s \\
\hline
\end{tabularx}
\caption{\textbf{Training and inference runtime comparison.} QCML required more training time but had the shortest measured inference time in this benchmark.}
\label{tab:runtime_performance}
\end{table}

In this study we use LST cutoff of 12 to identify LST+ cells. This threshold is based on previously published analytical validation data of the metastatic breast cancer platform \cite{RN2}, however, QCML achieved the highest mean balanced accuracy at each tested LST threshold as shown in \autoref{tab:LST_performance}.

\begin{table}[H]
\centering
\scriptsize
\renewcommand{\arraystretch}{1.2}
\begin{tabularx}{0.7\textwidth}{|l|Y|Y|Y|}
\hline
\multicolumn{4}{|c|}{\textbf{Balanced Accuracy : Fold validation}} \\
\hline
\diagbox{\textbf{Model}}{\textbf{Threshold}} & \textbf{LST=10} & \textbf{LST=12} & \textbf{LST=14} \\
\hline
\textbf{QCML} & \textbf{71.6\%} & \textbf{66.9\%} & \textbf{65.1\%} \\
\hline
\textbf{TabPFN} & 67.0\% & 65.5\% & 64.2\% \\
\hline
\textbf{CatBoost} & 62.7\% & 64.2\% & 64.6\% \\
\hline
\textbf{RBF SVM} & 63.0\% & 64.4\% & 64.9\% \\
\hline
\textbf{Elastic Net} & 62.0\% & 63.4\% & 62.9\% \\
\hline
\textbf{Linear SVM} & 64.3\% & 63.1\% & 64.8\% \\
\hline
\textbf{XGBoost} & 62.3\% & 61.2\% & 62.3\% \\
\hline
\textbf{MLP} & 66.4\% & 64.5\% & 60.6\% \\
\hline
\textbf{AdaBoost} & 64.6\% & 66.1\% & 64.4\% \\
\hline
\textbf{Nearest Neighbors 5} & 63.7\% & 62.6\% & 62.8\% \\
\hline
\textbf{Random Forest} & 60.3\% & 62.7\% & 63.6\% \\
\hline
\textbf{Nearest Neighbors 32} & 53.3\% & 56.3\% & 63.6\% \\
\hline
\textbf{Linear Regression} & 57.3\% & 56.3\% & 56.8\% \\
\hline
\end{tabularx}
\caption{\textbf{Regression model performance at different LST thresholds} Showing the fold-validation balanced accuracy across varying LST thresholds of 10, 12, 14. QCML achieved the highest mean balanced accuracy at each tested threshold.}
\label{tab:LST_performance}
\end{table}

\subsection{Gradient-based Feature Importance}
\begin{figure}[H]
    \centering
    \begin{minipage}{0.48\textwidth}
        \centering
        \includegraphics[trim=0 60 10 60, clip, width=\linewidth]{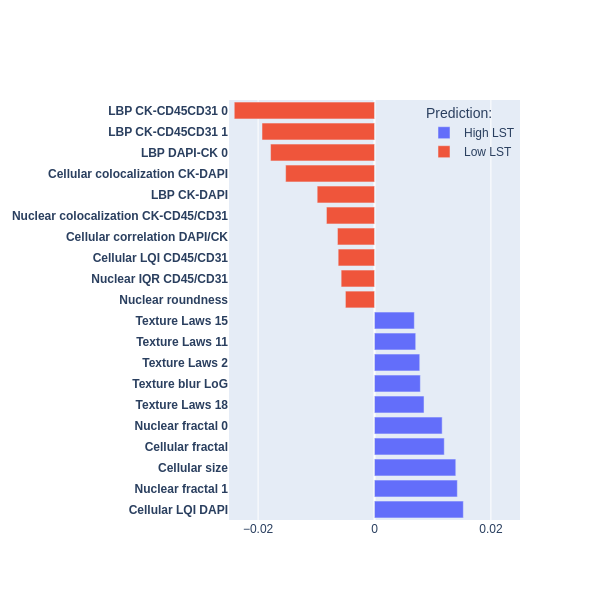}
        \label{fig:feature_importance}  
    \end{minipage}
    \hspace{1mm}
    \begin{minipage}{0.47\textwidth}
        \centering
        \renewcommand{\arraystretch}{1}
        \scriptsize
        \begin{tabular}{|l|c|c|}
        \hline
        \textbf{Features} & \textbf{Importance} & \textbf{Prediction} \\
        \hline
        LBP CK-CD45CD31 0 & 0.024 & Low LST \\
        \hline
        LBP CK-CD45CD31 1 & 0.019 & Low LST \\
        \hline
        LBP DAPI-CK 0 & 0.018 & Low LST \\
        \hline
        Cellular colocalization CK-DAPI & 0.015 & Low LST \\
        \hline
        Cellular LQI DAPI & 0.015 & High LST \\
        \hline
        Nuclear fractal 1 & 0.014 & High LST \\
        \hline
        Cellular size & 0.014 & High LST \\
        \hline
        Cellular fractal & 0.012 & High LST \\
        \hline
        Nuclear fractal 0 & 0.012 & High LST \\
        \hline
        LBP DAPI-CK 1 & 0.010 & Low LST \\
        \hline
        \end{tabular}
        \label{tab:feature_importance}
    \end{minipage}
    \caption{\textbf{QCML gradient-based feature importance.} \textbf{Left:} Top morphological and protein intensity features associated with high LST (blue) and low LST (red) predictions. \textbf{Right:} Top feature names based on absolute importance,  absolute importance score values, and model predictions (i.e., High LST or Low LST).}
    \label{fig:feature_importance_combined}
\end{figure}

QCML has several ways of identifying important features, one of them being the gradient-based feature importance which is based on the impact of changes in input features on the final output of the model. We use this approach to rank CTCs morphology features and protein expressions which were considered to be of high importance for the prediction. As shown in \autoref{fig:feature_importance_combined}, the top 10 CTCs morphology and protein content features that the QCML classifier used as predictor of LST are measures of: 

\begin{enumerate}[label=(\roman*)]
  \item Protein correlation and colocalization between channels such as Cross-channel (LBP) for CK vs DAPI and CK vs CD45/CD31 and cellular colocalization between CK-DAPI;
  \item Intensity features of the DAPI cellular mask;
  \item Cellular size;
  \item Fractal features.
\end{enumerate}

1) Cross-channel LBP and Cell colocalization features in the QCML model were predictors of low chromosomal instability which is in line with those features' ability to measure cellular remodeling and spatial discordance between nuclear and cytoplasmic structures \cite{RN62}. 

2) Interestingly, QCML weighed the signal of the lower DAPI quartile intensity from the cellular mask which indicates the cytoplasmic signal as one of the most important features in predicting higher levels of chromosomal instability (LST+). DAPI stains chromatin and double-stranded DNA which in normal cells typically resides in the nucleus and, for this reason, does not normally stain the cytoplasm \cite{RN69}. However, QCML data make sense with the underlying biology of cancer cells, as abnormal DAPI signals can appear in the cytoplasmic region due to the presence of micronuclei or nuclear envelope rupture, which typically occurs concomitantly with chromosomal instability \cite{RN64,RN76,RN77,RN83}, thus supporting the results of QCML. 

3) QCML identified cellular size, which is biologically relevant, as the presence of cytoplasmic micronuclei and multinucleation is permitted by larger-sized cells and is expected to occur in genomically unstable cells \cite{RN86,RN87}. Furthermore, cell size as a predictor of increased chromosomal instability is consistent with previous reports showing that CTC with larger metastatic breast cancer size were associated with the worst patient outcomes \cite{RN70,RN71,RN84,RN85} and for this reason larger cells are expected to be more likely to have increased chromosomal instability. 

4) Fractal features are approximations of irregularity and complexity in cellular and nuclear shape, suggesting abnormal nuclear contours that are expected to occur during chromosomal instability \cite{RN80,RN83}.

Within the present imaging and feature-extraction workflow, the texture cross-channel measures of colocalization identified by QCML suggest that CTCs with chromosomal instability may be more likely to have poorly aligned nuclear, subcellular, and cytoskeletal textures. Whether these associations and their relative feature importance are preserved across different imaging and staining conditions remains to be established.

\vspace{-.2in}

\subsection{Visualizing CTCs with QCML distance}

As mentioned in \Cref{sec:qcml}, QCML represents each observation as a quantum state. This allows one to have a natural notion of proximity between observations, since proximity between quantum states can be defined as {\em quantum fidelity} \cite[III.9]{nielsen00}
\[
f(\psi_1, \psi_2) = \lvert \braket{\psi_1}{\psi_2} \rvert ^ 2,
\]
which can be interpreted as the probability of identifying the state $\psi_1$ with the state $\psi_2$, when performing a quantum measurement designed to test whether a given quantum state is equal to $\psi_2$ (or vice versa). In the context of QCML, this type of proximity can be used to define a similarity measure on the data. Given the mapping from an observation to quantum state $\mathbf{x}_t \rightarrow \ket{\psi_t}$, we can define the {\em QCML distance} between two data points $\mathbf{x}_t, \mathbf{x}_{t'}$ as
\begin{equation}
\label{eqn:QCML_distance}
d_Q(\mathbf{x}_t, \mathbf{x}_{t'}) = 1 - f(\psi_t, \psi_{t'}) = 1 - \lvert \braket{\psi_t}{\psi_{t'}} \rvert ^ 2.
\end{equation}

Note that in contrast to the standard Euclidean distance between two data points, the QCML distance is a type of supervised similarity measure, since the representation of the data in quantum states $\psi_t$ has been optimized using the training targets.

\begin{figure}[H]
    \centering
    \includegraphics[width=0.8\linewidth]{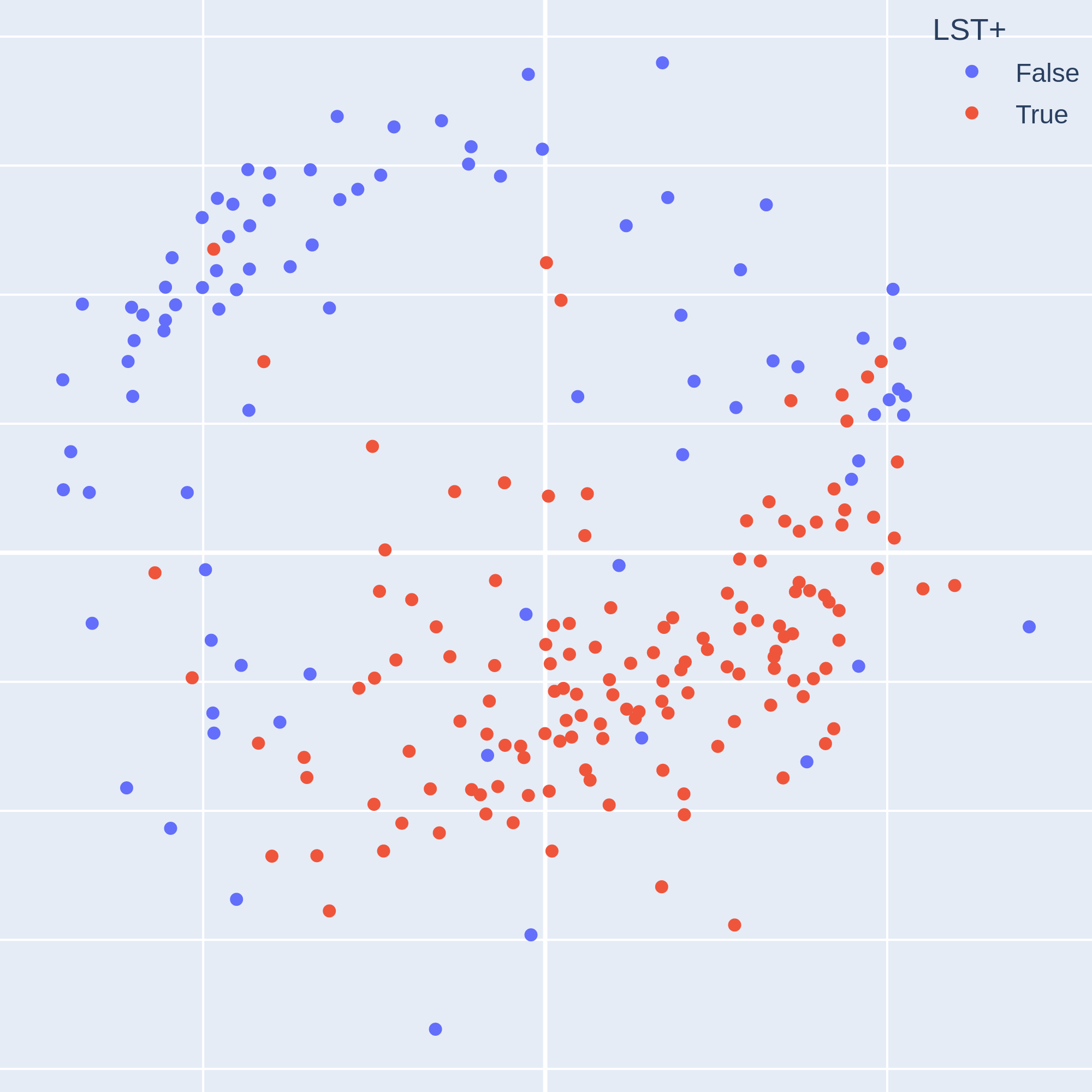}
    \caption{\textbf{Multi-dimensional scaling visualization based on a distance matrix of CTCs.} Red color represents LST+ CTCs, while blue color represents stable CTCs. Using QCML distance one can achieve a visual separation between LST+ and stable CTCs. }
    \label{fig:MDS}
\end{figure}

Given a distance matrix, we can visualize the observations in a two-dimensional space using multidimensional scaling (MDS). This is a common dimensionality reduction technique that can be used to visualize high-dimensional data in two dimensions. In a nutshell, MDS finds a mapping of the high-dimensional data into two dimensions that minimizes the matrix norm of the difference between the distance matrix of the original data and that of the two-dimensional transformation. Using MDS we plot the high-dimensional CTCs data in two dimensions, as shown in \autoref{fig:MDS}.

Although the plots in two-dimensional MDS target space do not offer a completely faithful representation of the distances between CTCs, they help qualitatively illustrate the ability of QCML distance to find a separation between genomically unstable and stable CTCs.

\section{Discussion}
\label{sec:conclusion}

Here, we applied Quantum Cognition Machine Learning (QCML) to digital pathology-derived morphological features and protein expression levels from CTCs, to enable prediction of chromosomal instability. The QCML-based pLST model achieved the highest mean balanced accuracy among the evaluated regression models. The ultimate objective of deploying a morphology prediction of LST algorithm in liquid biopsy CTC assays is to enable real-time detection of CTCs with elevated level of chromosomal instability and high metastatic potential. As a result, pLST detection can reduce diagnostic latency and circumvent the long turnaround times associated with whole genome sequencing in clinical workflows. 

Chromosomal instability is a key molecular driver of tumor heterogeneity, which in turn supports the high plasticity and evolutionary adaptability of cancer cells in response to environmental pressures, ultimately enabling metastasis \cite{RN6,RN8,RN14,RN16,RN23,RN24,RN26,RN34,RN49}. Therefore, the detection of cancer cells through preconceived expert knowledge of their expected biological phenotype may be a limitation, as it may not account for tumor evolution and the emergence of new CTC phenotypes \cite{RN18,RN40,RN42}. To address this challenge, the present study employed a previously published\cite{RN2,RN56}, systematic and quantifiable approach to nuclear and cellular segmentation using digital pathology to extract the broadest set of morphological, texture and intensity-based characteristics~\cite{RN2,RN56}. This approach provides an optimal foundation for the application of advanced machine learning models, including QCML. Looking ahead, future research should explore quantum-assisted feature mapping to identify complex and subtle patterns directly from raw image pixel data to see whether such an approach may potentially surpass the performance of biologically informed feature extraction. This direction may further enhance the ability to detect chromosomal instability in CTCs without relying on guided feature extraction.

One of the key challenges in applying machine learning to single cell diagnostics is the inherent biological complexity, which becomes especially problematic when the number of features (e.g., genes, or digital pathology features) far exceeds the number of samples \cite{RN67,RN74}. This imbalance, common in genomics, proteomics, and digital pathology, can lead to overfitting and poor generalization \cite{RN67}. Classical statistical methods, such as Bayesian inference, often require data volumes that grow exponentially with the number of features, making them impractical in such high-dimensional biological settings \cite{RN27,RN29,RN55}. To address this, modern approaches in genomics and digital pathology typically reduce dimensionality by selecting curated ``signatures'' based on preconceived biological knowledge. Rather than relying on statistical associations on all available features, these curated features reflect a mechanistic understanding of biological systems. The interpretability of the features typically reduces the overfitting to the training data and improves the robustness and reproducibility of the model \cite{RN27,RN29,RN55}. In other words, we apply our human cognition to design predictive models that make sense and draw conclusions ignoring irrelevant information. However, such feature engineering is a way to offset machine learning limitations by leveraging human intervention and its understanding of the biological problem \cite{RN75,RN67}. 

It has been proposed that machine learning algorithms should learn representations of the data by disentangling explanatory factors, mimicking human cognition in understanding disease mechanisms \cite{RN75,RN67}. In a similar attitude, more advanced and recent approaches leverage principles from quantum theory to address high-dimensional data representation by simulating cognition \cite{CandeloriEtAl, BlackrockBondSimilarity}. By adopting the formalism of quantum probabilities, particularly the uncertainty principle, data can be encoded as vectors within a Hilbert space, where no state corresponds to an exact position of the features configuration. Therefore, through a simulation of quantum principles using classical computers, we enable an intrinsic reduction in feature representation, offering a novel way to manage complexity and dimensionality in biomedical data analysis. 

Following these principles, in the present study, QCML was applied to abstract out the features that are the most fundamental to estimate the intrinsic dimensions of the CTC morphology data. By doing this, and without human curation, QCML learned a gradient of feature importance that was found \textit{posthoc} to be biologically and mechanistically involved with chromosomal instability in cancer cells. QCML's prediction of chromosomal instability classification abstracted a model for instability-driven morphological shifts where CTCs are larger in cellular size with higher spatial discordance between nuclear and cytoplasmic structures. The texture cross-channel measures of colocalization identified by QCML suggest that CTCs with chromosomal instability may be more likely to have poorly aligned nuclear, subcellular and cytoskeletal textures. Those indicate structural rearrangements and nuclear pleomorphism which have been previously linked to genomically unstable tumor cells \cite{RN65}. In addition, the morphological manifestation of chromosomal instability, which can be perceived as lower cellular integrity, has been shown to provide functions that could ultimately be evolutionary advantageous for cancer \cite{RN64}. QCML findings were also corroborated by the evidence that the cellular localization of DAPI intensity and size is important for the prediction of chromosomal instability. In cancer cells, the distinction between nuclear and cellular (cytoplasmic) DAPI expression becomes especially important, as abnormalities in DAPI localization and intensity (e.g. small, round DAPI-positive bodies in cytoplasm) can reveal hallmarks of chromosomal instability, and architecture defects \cite{RN64,RN69,RN76,RN77,RN83}. 

Future studies will be required to validate these findings and establish whether the presence of CTC with predicted chromosomal instability classified by QCML can predict patient survival with better performance compared to conventional methods.

\bibliography{references}

\section*{Author contributions}
G.D.C. designed the study, and drafted the manuscript. V.K. designed and performed the QCML analysis. E.T.L. extracted the image features. A.G.A., J.R.B., L.C., N.H., K.M., R.S., H.S., D.V., M.T.W., R.J.W., M.X. contributed to the research, providing feedback and reviewing the manuscript.

\section*{Data Availability}
The datasets generated during and/or analyzed during the current study are not publicly available due to privacy and ethical restrictions but are available from the corresponding authors on reasonable request and with permission of Epic Sciences.

\section*{Competing Interests}
The authors declare that they have no competing interests. Research support for this study was funded by Epic Sciences and Qognitive.

\end{document}